\documentclass[onecollarge,natbib]{svjour2}
\bibpunct{[}{]}{,}{n}{}{,} 
\smartqed  
\usepackage{graphicx}
\usepackage{amsmath}
\usepackage{amsfonts}
\usepackage{amssymb}
\newcommand{\hatbf}[1]{\hat{{\mathbf #1}}}
\journalname{Few-Body Systems (APFB2011)}
\begin{document}

\title{\boldmath
Scattering of a spin-$\frac{1}{2}$ particle off a spin-$0$ target in a simple
three-dimensional basis}
\subtitle{}

\titlerunning{Scattering of a spin-$\frac{1}{2}$ particle off a spin-$0$ target in a simple 3D basis}        

\author{Imam Fachruddin         \and
        Agus Salam 
}

\authorrunning{I. Fachruddin \& A. Salam} 

\institute{I. Fachruddin \at
              Departemen Fisika, Universitas Indonesia, Depok 16424, Indonesia \\
              \email{imamf@fisika.ui.ac.id}           
           \and
           A. Salam \at
              Departemen Fisika, Universitas Indonesia, Depok 16424, Indonesia
}

\date{Received: date / Accepted: date}

\maketitle

\begin{abstract}
Scattering of a spin-$\frac{1}{2}$ particle off a spin-$0$ target is formulated based on a simple 
three-dimensional momentum-spin basis. The azimuthal behaviour of both the potential and the
$T$-matrix elements leads to a set of integral equations for the $T$-matrix elements in two
variables only, namely the momentum's magnitude and the scattering angle. Some symmetry relations
for the potential and the $T$-matrix elements reduce the number of the integral equations to be
solved by a factor of one half. A complete list of the spin observables in terms of the
two-dimensional $T$-matrix elements is presented. 
\end{abstract}

\section{Introduction}
\label{intro}
The standard partial-wave (PW) technique has been successfully used to calculate few
particle system, see for example Ref.~\cite{rep} for 3 nucleon systems. But as energy
increases PW calculations become more tedious, since many more higher angular momentum
states have to be included. As an alternative to the PW technique one can use a
so called three-dimensional (3D) technique, the basic idea of which is not to expand
the basis states into partial waves. Many works based on this idea for various systems 
have been carried out. See for example Refs.~\cite{thomas, liu} for 2 and 3 boson
scattering, Refs.~\cite{holz,rice,nn3d,golak} for 2 nucleon scattering,
Ref.~\cite{gloeckle} for 3 nucleon scattering, and Ref.~\cite{irga} for scattering of a
spin-$\frac{1}{2}$ particle and a spin-$0$. 

We take as our system the same system being considered in Ref.~\cite{irga}, but we define
a different basis. In fact our basis states are simpler than those used in
Ref.~\cite{irga}. We, nevertheless, end up with some integral equations, the solution of
which can be directly used to calculate observables, which is not the case in
Ref.~\cite{irga}. 

We show in Section \ref{form:1} the basis states and the derivation of the integral
equations for the $T$-matrix elements based on these basis states. In Section
\ref{form:2} we show a complete list of spin observables being connected directly to the
$T$-matrix elements, which are the solution of the obtained integral equations. We sumarize in
Section \ref{summary}.

\section{Formulation}
\label{formulation}
\subsection{T-matrix elements}
\label{form:1}
We define our basis state $\bigl|\mathbf{p}\lambda\bigr>$ as a direct product of the free 
state $\bigl|\mathbf{p}\bigr>$ and the spin $s = \frac{1}{2}$ state
$\bigl|\hatbf{z}\lambda\bigr>$: 
\begin{equation}
\bigl|\mathbf{p}\lambda\bigr> \equiv \bigl|\mathbf{p}\bigr> \bigl|\hatbf{z}\lambda\bigr>
\, . \label{bs}
\end{equation}
Here $\mathbf{p}$ and $s = \frac{1}{2}$ are the relative momentum and the total spin of
the two particles, with the spin being quantized along the z-axis. Let us just
call $\bigl|\mathbf{p}\lambda\bigr>$ the 3D basis states. 
Based on the 3D basis states the potential $V$- and $T$-matrix elements are defined as
\begin{eqnarray}
V_{\lambda' \lambda}(\mathbf{p}',\mathbf{p}) 
& \equiv & \bigl<\mathbf{p}'\lambda'\bigr| V \bigl|\mathbf{p}\lambda\bigr> \label{3dv}\\ 
T_{\lambda' \lambda}(\mathbf{p}',\mathbf{p}) 
& \equiv & \bigl<\mathbf{p}'\lambda'\bigr| T \bigl|\mathbf{p}\lambda\bigr> \, . 
\end{eqnarray}
These $T$-matrix elements $T_{\lambda' \lambda}(\mathbf{p}',\mathbf{p})$
obey the following Lippmann-Schwinger equation
\begin{equation}
T_{\lambda' \lambda}(\mathbf{p}',\mathbf{p}) 
= V_{\lambda' \lambda}(\mathbf{p}',\mathbf{p}) + \sum_{\lambda'' =
-\frac{1}{2}}^{\frac{1}{2}} \int d\mathbf{p}'' V_{\lambda'
\lambda''}(\mathbf{p}',\mathbf{p}'') G_0^+(E_p) 
T_{\lambda'' \lambda}(\mathbf{p}'',\mathbf{p}) \, ,
\end{equation}
with
\begin{equation}
G_0^+(E_p) = \lim _{\epsilon \rightarrow 0} \frac{1}{E_p + i \epsilon - E_{p''}}
\, , \quad E_p = \frac{p^{2}}{2\mu} \, .
\end{equation}

For the system being considered the interaction takes a general structure given as 
\begin{equation}
V\left( \mathbf{p}', \mathbf{p}\right) 
\equiv \bigl<\mathbf{p}'\bigr| V \bigl|\mathbf{p}\bigr> 
= f_{0}\left(p',p,\hatbf{p}' \cdot \hatbf{p} \right) 
+ f_{1}\left(p',p,\hatbf{p}' \cdot \hatbf{p} \right) 
\bigl( \mathbf{s} \cdot \hatbf{p}' \bigr) \bigl( \mathbf{s} \cdot \hatbf{p} \bigr) \, ,
\label{genpot}
\end{equation}
with $\mathbf{s} = \frac{1}{2} \boldsymbol{\sigma}$, $\boldsymbol{\sigma}$ the Pauli spin
operator, $f_{i}\left(p',p,\hatbf{p}' \cdot \hatbf{p} \right) , \, (i = 0,1)$
spin-independent functions. Assuming as well time-reversal invariance, this requires 
\begin{equation}
f_{i}\left(p',p,\hatbf{p}' \cdot \hatbf{p} \right) =
f_{i}\left(p,p',\hatbf{p} \cdot \hatbf{p}' \right) \, , \, (i = 0,1) \, . 
\end{equation}
Inserting Eq.~(\ref{genpot}) into Eq.~(\ref{3dv}) we get the potential matrix elements
$V_{\lambda' \lambda}(\mathbf{p}',\mathbf{p})$ as 
\begin{eqnarray}
V_{\lambda' \lambda}(\mathbf{p}',\mathbf{p}) 
& = & \delta_{\lambda' \lambda} \Bigl[ f_{0}\left(p',p,\hatbf{p}' \cdot \hatbf{p}
\right) 
+ \frac{1}{4} f_{1}\left(p',p,\hatbf{p}' \cdot \hatbf{p} \right) \left\lbrace \cos \theta'
\cos \theta 
+ e^{- 2 i \lambda (\phi' - \phi)} \sin \theta' \sin \theta \right\rbrace \Bigr] 
\cr 
&& + \delta_{\lambda', -\lambda} f_{1}\left(p',p,\hatbf{p}' \cdot \hatbf{p} \right) 
\frac{\lambda}{2} e^{ 2 i \lambda \phi'} \left\lbrace \sin \theta' \cos \theta 
- e^{- 2 i \lambda (\phi' - \phi)} \cos \theta' \sin \theta \right\rbrace \, . 
\end{eqnarray}

In the case, where $\hatbf{p} = \hatbf{z}$, the azimuthal behaviour of $V_{\lambda'
\lambda}(\mathbf{p}',\mathbf{p})$ shows up as 
\begin{equation}
V_{\lambda' \lambda}(\mathbf{p}',p\hatbf{z}) 
= e^{-i (\lambda' - \lambda) \phi'} V_{\lambda' \lambda}(p',p,\theta') \, , \label{aziv}
\end{equation}
with 
\begin{equation}
V_{\lambda' \lambda}(p',p,\theta') 
= \delta_{\lambda' \lambda} \left\lbrace f_{0}\left(p',p,\cos \theta' \right) 
+ \frac{1}{4} f_{1}\left(p',p,\cos \theta' \right) \cos \theta' \right\rbrace 
+ \delta_{\lambda', -\lambda} \frac{\lambda}{2} f_{1}\left(p',p,\cos \theta'
\right) \sin \theta' \, . \label{vnazim}
\end{equation}
It can be shown that the azimuthal behaviour of $V_{\lambda'
\lambda}(\mathbf{p}',p\hatbf{z})$ given in Eq.~(\ref{aziv}) also applies to $T_{\lambda'
\lambda}(\mathbf{p}',p\hatbf{z})$, thus, 
\begin{equation}
T_{\lambda' \lambda}(\mathbf{p}',p\hatbf{z}) 
= e^{-i (\lambda' - \lambda) \phi'} T_{\lambda' \lambda}(p',p,\theta') \, . \label{azit}
\end{equation}
The $T$-matrix elements $T_{\lambda' \lambda}(p',p,\theta')$ defined in Eq.~(\ref{azit})
obey the following integral equation: 
\begin{eqnarray}
T_{\lambda' \lambda}(p',p,\theta') 
& = & V_{\lambda' \lambda}(p',p,\theta') \cr 
&& + 2 \mu \lim_{\epsilon \rightarrow 0} \sum_{\lambda'' =
-\frac{1}{2}}^{\frac{1}{2}} 
\int\limits_{0}^{\infty} dp'' \frac{p''^2}{p^2 + i\epsilon - p''^2} \int\limits_{-1}^{1}
d\cos \theta'' V_{\lambda' \lambda''}^{\lambda}(p',p'',\theta',\theta'') 
T_{\lambda'' \lambda}(p'',p,\theta'') \, , \qquad \label{2dlse}
\end{eqnarray}
with 
\begin{equation}
V_{\lambda' \lambda''}^{\lambda}(p',p'',\theta',\theta'') 
\equiv \int\limits_{0}^{2 \pi} d\phi'' V_{\lambda' \lambda''}(\mathbf{p}',\mathbf{p}'')
e^{ i (\lambda' \phi' - \lambda'' \phi'')} e^{-i \lambda (\phi' - \phi'')} \, . 
\end{equation}

Some symmetry relations for $V_{\lambda' \lambda}(p',p,\theta')$ and
$V_{\lambda' \lambda''}^{\lambda}(p',p'',\theta',\theta'')$ are obtained as 
\begin{eqnarray}
V_{\lambda' \lambda}(p',p,\theta') 
& = & (-)^{\lambda' - \lambda} V_{-\lambda',-\lambda}(p',p,\theta') \label{symv}\\
V_{\lambda' \lambda''}^{\lambda}(p',p'',\theta',\theta'') 
& = & (-)^{\lambda' - \lambda''}
V_{-\lambda',-\lambda''}^{-\lambda}(p',p'',\theta',\theta'') \, . \label{symvkern}
\end{eqnarray}
Applying these symmetry relations, Eqs.~(\ref{symv}) and (\ref{symvkern}), in the
integral equation given in Eq.~(\ref{2dlse}) leads to similar symmetry behavior for
$T_{\lambda' \lambda}(p',p,\theta')$, that is
\begin{equation}
T_{\lambda' \lambda}(p',p,\theta') 
= (-)^{\lambda'-\lambda} T_{-\lambda',-\lambda}(p',p,\theta') \, . 
\end{equation}
We need, hence, to solve only 1 set of 2 coupled equations of Eq.~(\ref{2dlse}), namely 
those for $T_{\frac{1}{2} \frac{1}{2}}(p',p,\theta')$ and $T_{-\frac{1}{2}
\frac{1}{2}}(p',p,\theta')$. 

\subsection{Spin observables}
\label{form:2}
In this section we show some spin observables for the case, where the spin-$\frac{1}{2}$
particle plays as the projectile (particle 1) and the spin-$0$ particle as the target
(particle 2). These observables are spin-averaged differential cross section,
polarization, analyzing power, and depolarization tensor. These observables can be
calculated directly from the $T$-matrix elements $T_{\frac{1}{2}
\frac{1}{2}}(p',p,\theta')$ and $T_{-\frac{1}{2} \frac{1}{2}}(p',p,\theta')$ as: 

\paragraph{spin-averaged differential cross section:}
\begin{equation}
I_0 \equiv \overline{\frac{d\sigma}{d\hatbf{p}'}} 
= (4 \pi^2 \mu)^2 \left\lbrace \left| T_{\frac{1}{2} \frac{1}{2}}(p,p,\theta')
\right|^2 + \left| T_{-\frac{1}{2} \frac{1}{2}}(p,p,\theta') \right|^2 \right\rbrace 
\end{equation}

\paragraph{polarization $P_y$ and analyzing power $A_y$:}
\begin{equation}
P_y = A_y = \frac{2}{I_0} (4 \pi^2 \mu)^2 
Im \left\lbrace T_{\frac{1}{2} \frac{1}{2}}^{\ast}(p,p,\theta') T_{-\frac{1}{2}
\frac{1}{2}}(p,p,\theta') \right\rbrace 
\end{equation}

\paragraph{depolarization tensor:}
\begin{eqnarray}
D_{x' x} = D_{z' z} 
& = & \frac{1}{I_0} (4 \pi^2 \mu)^2 
\biggl[ \left\lbrace \left| T_{\frac{1}{2} \frac{1}{2}}(p,p,\theta') \right|^2 
- \left| T_{-\frac{1}{2} \frac{1}{2}}(p,p,\theta') \right|^2 \right\rbrace \cos
\theta_{lab} \cr
&& \qquad \qquad \quad 
+ 2 Re \left\lbrace T_{\frac{1}{2} \frac{1}{2}}^{\ast}(p,p,\theta') 
T_{-\frac{1}{2} \frac{1}{2}}(p,p,\theta') \right\rbrace \sin \theta_{lab} \biggr] \\
D_{z' x} = - D_{x' z} 
& = & \frac{1}{I_0} (4 \pi^2 \mu)^2 
\biggl[ \left\lbrace \left| T_{\frac{1}{2} \frac{1}{2}}(p,p,\theta') \right|^2 
- \left| T_{-\frac{1}{2} \frac{1}{2}}(p,p,\theta') \right|^2 \right\rbrace \sin
\theta_{lab} \cr
&& \qquad \qquad \quad 
- 2 Re \left\lbrace T_{\frac{1}{2} \frac{1}{2}}^{\ast}(p,p,\theta') 
T_{-\frac{1}{2} \frac{1}{2}}(p,p,\theta') \right\rbrace \cos \theta_{lab} \biggr] \\ 
D_{y y} 
& = & 1 
\end{eqnarray}
with 
\begin{equation}
\tan \theta_{lab} = \left( \frac{m_2}{m_1 \sec \theta' + m_2} \right) \tan \theta' \, . 
\end{equation}

\section{Summary}
\label{summary}
We have formulated the Lippmann-Schwinger equations based on a simple 3D basis for a system of a
spin-$\frac{1}{2}$ particle and a spin-$0$ particle. Especially for higher energies this can be use
to calculate, for example, kaon-nucleon scattering. The resulting final equations are integral
equations for some $T$-matrix elements, which depend on the momentum's magnitude and the scattering
angle. Symmetry relations allow us to solve instead of two only 1 set of two of these equations. We
derive a complete list of the spin observables for scattering of a spin-$\frac{1}{2}$ particle off a
spin-$0$ target, showing that these observables can be calculated directly from the solutions of the
integral equations.


\begin{thebibliography}{8}
\bibitem{rep} Gl\"ockle, W., et.al.: Phys. Rep. 274, 107 (1996)
\bibitem{thomas} Elster, Ch., Thomas, J.H., Gl\"ockle, W.: Few-Body Systems 24, 55 (1998)
\bibitem{liu} Liu, H., Elster, Ch., Gl\"ockle, W.: Phys. Rev. C 72, 054003 (2005)
\bibitem{holz} Holz, J., Gl\"ockle, W.: Phys. Rev. C 37, 1386 (1988)
\bibitem{rice} Rice, R.A., Kim, Y.E.: Few-Body Systems 14, 127 (1993)
\bibitem{nn3d} Fachruddin, I., Elster, Ch., Gl\"ockle, W.: Phys. Rev. C 62, 044002 (2000)
\bibitem{golak} Golak, J., et.al.: Phys. Rev. C 81, 034006 (2010)
\bibitem{gloeckle} Gl\"ockle, W., et.al.: Eur. Phys. J. A 43, 339 (2010)
\bibitem{irga} Abdulrahman, I., Fachruddin, I.: Mod. Phys. Lett. A 24, 843 (2009)

\end{thebibliography}
\end{document}